\documentclass[aps,prl,amsmath,twocolumn,amssymb,floatfixng,showpacs,
superscriptaddress,footinbib]{revtex4-1}
\pdfoutput=1
\usepackage[dvips]{graphics}
\usepackage{bm}
\usepackage{float}
\usepackage{epsfig}
\usepackage{enumerate}
\usepackage{amsmath}
\usepackage{color}
\usepackage{graphicx}
\usepackage[colorlinks=true,linktoc=page,linkcolor=red,citecolor=blue]{hyperref}

\newcommand\bea{\begin{eqnarray}}
\newcommand\eea{\end{eqnarray}}
\newcommand\beq{\begin{equation}}  
\newcommand\eeq{\end{equation}}

\begin{document}
	
	\title{\titlename}
	\date{\today}
	\title{Light-driven Lifshitz transitions in non-Hermitian multi-Weyl semimetals}
	\author{Debashree Chowdhury}
	\email{debashreephys@gmail.com}
	\affiliation{Centre of Nanotechnology, Indian Institute of Technology Roorkee, Roorkee, Uttarakhand-247667}
	\author{Ayan Banerjee}
	\email{ayanbanerjee@iisc.ac.in}
	\affiliation{Solid State and Structural Chemistry Unit, Indian Institute of Science, Bangalore 560012, India}
	\author{Awadhesh Narayan}
	\email{awadhesh@iisc.ac.in}
	\affiliation{Solid State and Structural Chemistry Unit, Indian Institute of Science, Bangalore 560012, India}
	
	\date{\today}
	\begin{abstract}
		
		Non-Hermitian topological systems are the newest additions to the growing field of topological matter. In this work, we report of the light-driven exceptional physics in a multi-Weyl semi-metal. The driving is not only a key ingredient to control the position of the exceptional contours (ECs), light also has the ability to generate new ECs. Interestingly, we also demonstrate topological charge distribution and Lifshitz transition, which are controllable by the driving field in such generated ECs. Our findings present a promising platform for the manipulation and control over exceptional physics in non-Hermitian topological matter.
		
	\end{abstract}
	
	\maketitle
	
	Topology plays a pivotal role in the study of condensed matter systems -- with wide focus on topological insulators and superconductors in the last decade~\cite{Hasan,Vafek,Kane}. The new feather in the cap is the study of topological semimetals (TSMs)~\cite{Burkov}, among which Weyl semimetals (WSMs)~\cite{Armitage} have attracted a great recent interest. In contrast to a topological insulator, WSMs are gapless in the bulk and break either or both time reversal and inversion symmetries. The WSM spectra is linear near Weyl points, which usually come in pairs and act as sources and sinks of the Berry curvature with monopole charge $\pm 1.$ Furthermore, WSMs with quadratic, cubic (or even higher order) dispersion have also been proposed~\cite{Fang,Xu1,Huang,Zhang,Gupta}. These are coined as double, triple (or, in general, multi) WSMs. Besides having a non-linear dispersion relation, these WSMs are unique in the sense that they are protected by rotation symmetries of different point groups, specifically the $n$-fold rotation symmetry $C_{n}$~\cite{Fang,Xu1,Huang,Zhang,Gupta,MW2,MW1,TW} (see supplementary material) and they have integer topological charges greater than unity. 
	
	In the last few years, the analysis of non-Hermitian (NH)
	gain and loss terms on different topological properties have captivated the research community~\cite{Ghatak,Bergholtz,Gong,Jean,Bandres,f14,f15,f16,f17,f18,f19,f20,Banerjee,Bliokh,Lee,He}. Many experimental efforts have strengthened the theoretical predictions in systems such as ultra-cold atoms~\cite{Goldman,Xu} and optics~\cite{Noh}. An unconventional feature that makes these systems special is their defectiveness, i.e., merging of eigenstates at some particular points, where the eigenenergies also coalesce~\cite{Luis}. These special degenerate points are called exceptional points (EPs)~\cite{Luis,Cerjan,Cerjan1,FM}. In addition to having a single EP, there may also arise exceptional surfaces with a collection of EPs, termed exceptional contours (ECs) in a multi-dimensional parameter space~\cite{Bergholtz,Cerjan1,Cerjan2}.

	Tuning properties of quantum systems with light has been a promising new frontier in the recent years~\cite{f1,f2,f3,f4,f5,f6,f7,f8,f9,f10,f11,f12,f13,f13a,f13b,f13c,f13d}. Interaction with topology has led to unveiling of numerous exciting phenomena -- for instance in conventional TSMs the light induces new phases~\cite{f2} and can cause profound changes in the Fermi surface topology, which is coined as the Lifshitz transition\cite{1,2,3,Beaulieu,f13e}. On the other hand, in linear WSMs circularly polarised light (CPL) can cause a tuning of the distance between the two Weyl points resulting in an anomalous Hall effect~\cite{Chen1,Chen2}. 
	
\begin{figure}
		\includegraphics[scale=0.5]{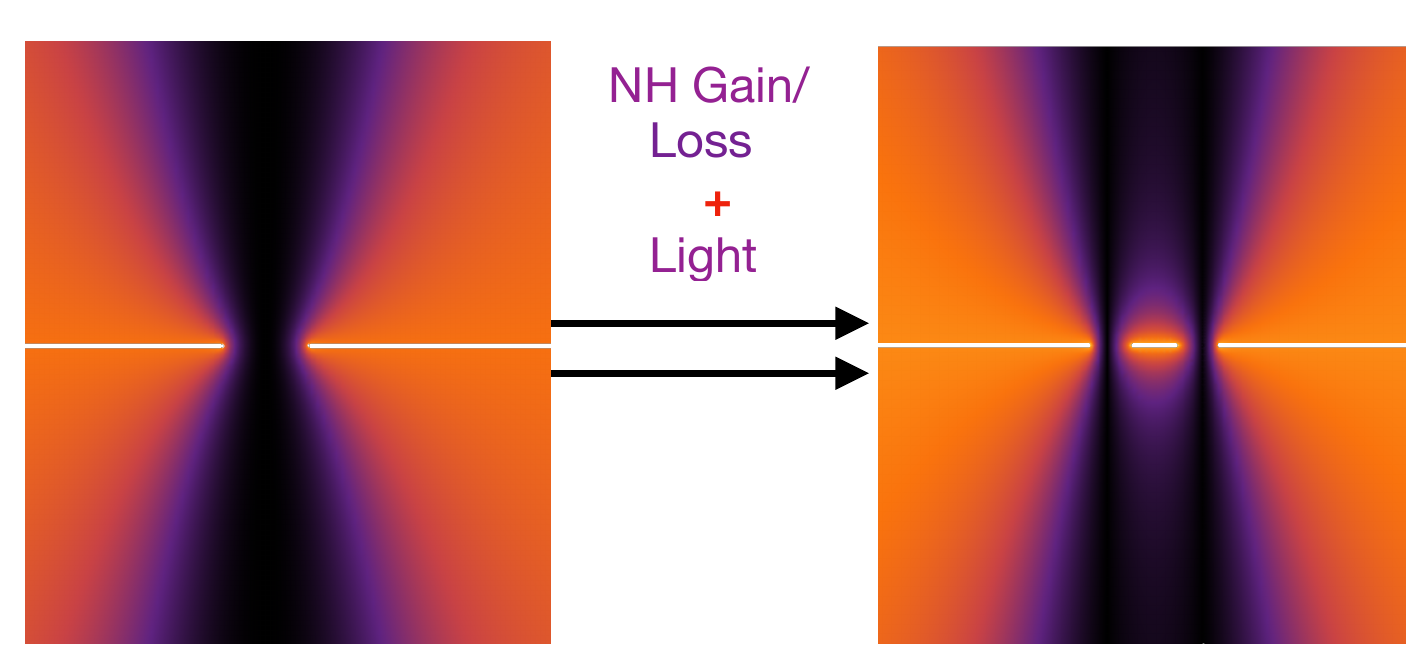}
		\caption{\textbf{Schematic of proposal for light-induced Lifshitz transition in a double Weyl semimetal.} A combination of NH terms and driving converts a single exceptional ring into two exceptional rings.}  \label{Schematic}
	\end{figure}

	In this letter, we have analyzed the ECs that appear in NH multi-WSMs in presence of CPL. We show that the driving can be used as a facile tool to generate ECs. In addition, besides controlling the position of the exceptional rings one can also generate new ECs. Furthermore, we study the charge division of the newly created ECs by means of NH generalization of the Berry curvature, and the concomitant Lifshitz transitions. In Fig.~\ref{Schematic}, we present a schematic for the Lifshitz transition of a double Weyl semimetal and the generation of new ECs in presence of both CPL and gain/loss parameter. Importantly, one can notice that the exceptional ring, which appears due to the gain/loss term in absence of light (left panel in Fig.~\ref{Schematic}), divides into two exceptional rings when light is switched on.
	
	We start with the analysis of effects of driving on the band structure of NH multi-WSMs. The low energy Hamiltonian of an $n$-th order multi-WSM in presence of the NH loss or gain term is~\cite{Cerjan1}
	
	\begin{equation}
	\label{Hmn}
	H^{\eta,n}(q)=\frac{1}{2m}\Big(q_{-}^{n}\sigma_{+}+q_{+}^{n}\sigma_{-}\Big)+\left(\eta v_z q_z+i\zeta\right)\sigma _z,
	\end{equation}
	
	where $\zeta$ is the gain or loss parameter. Here $q_{\pm} =q_{x}\pm i q_{y}$ and $\sigma_{\pm} =\sigma_{x}\pm i \sigma_{y}$, with $\sigma_{i}$ and $q_{i}$ ($i=x,y,z$) denoting the components of Pauli matrices and quasi-momentum respectively. Here, $m$ is the quasi-particle mass, $\eta=\pm 1$ and $v_{z}$ is the $z$ component of the Fermi velocity. The corresponding energy eigenvalues are in general complex. This complex spectra of the system causes some accidental degeneracies, where not only the eigen-energies coalesce, but also eigen-vectors merge -- these degeneracy points are coined as EPs. Our aim is to find out the exact positions of these EPs and the effect of these degenarecies on different physical properties. In case of a usual WSM, it is shown in the previous work~\cite{Cerjan} that adding a NH term to the Hamiltonian can convert the Weyl point into an exceptional ring, which carries the same charge as that of the original Weyl point~\cite{Cerjan1}. Furthermore, depending on the strength of the gain/loss term, it is also possible to merge two ECs of opposite charge and in the process of this merging, they annihilate and form a single contour~\cite{Cerjan1}.
	
	In order to investigate the physical properties of such a NH system under driving, let us include an optical field polarized in the $x-z$ plane of the form ${\cal E}(t)={\cal E}_{0}(\cos \Omega t,0,-\sin \Omega t)$, where ${\cal E}_{0}$ and $\Omega$ are the amplitude and frequency of the driving optical field. The minimal substitution leads to $\hbar q_{i} \rightarrow \hbar q_{i}+eA_{i}$, where $e$ is the electronic charge, $A$ is the vector potential with $\vec{A}(t+T)=\vec{A}(t)$, and $T=2\pi/ \Omega$ as the periodicity. Floquet theory is an elegant method to incorporate the periodic driving effects on the band structure. In Floquet formalism we consider a time dependent periodic Hamiltonian $H(t+T)=H(t)$, that exhibits quasiperiodic eigenspectra. Interestingly, non-Hermitian Floquet systems admit complex quasienergies with some periodicity leading to a point gap topology with non-trivial winding associated with quantized charge transport, which has no analog in conventional Hermitian physics~\cite{F}. Although Floquet theory is valid for all frequency regimes of the driving field, here we restrict ourselves to the high frequency (HF) regime, where the frequency of the optical field is much larger than the bandwidth of the system. This HF approximation has similarities with the rotating wave approximation. This eventually breaks the time reversal symmetry (TRS) and as a result a gap opening can be observed in several systems. Importantly, in charge $\pm 1$ WSMs the gap opening at the Weyl point is not possible with this HF laser light. Rather the optical field leads to a shift of the Weyl points of the TRS broken WSMs. For multi-WSMs, interestingly, we find a rather interesting role of light in presence of loss/gain terms. In the HF approximation, it is possible to easily calculate the effective time independent Hamiltonian using the Floquet-Magnus expansion~\cite{f1,f2,f6,He}. Thus we have (see supplementary material),
	\begin{align}\label{12}
	&H_{\mathrm{eff}}^{n,\eta}(q)=\frac{1}{2m}\Big((q_{-}^{n}+i \Delta_{}q_{-}^{n-1}) \sigma_{+}\nonumber\\&+(q_{+}^{n}-i \Delta_{}q_{+}^{n-1})\sigma_{-}\Big)+\left(\eta v_{z} q_{z}+i\zeta\right)\sigma _z,
	\end{align}
	
	where $\Delta_{} =\frac{nA_{0}^{2}v_{z}\eta}{2\hbar \Omega},$ where $A_{0}$ and $\Omega$ are the amplitude and frequency of the driving field respectively. 
	
	The corresponding energy eigenvalues are
	\begin{align}\label{13}
	&{\cal E}^{}_{m,\pm}=\nonumber\\&\pm\sqrt{\frac{(q_{x}^{2}+  q_{y}^{2})^{(n-1)} (q_{x}^{2} + (q_{y} - \Delta_{})^2) }{m^{2}}-(\zeta - 
		i q_{z} v_{z} \eta)^2}.
	\end{align}
	
	\begin{figure}
		\includegraphics[scale=0.138]{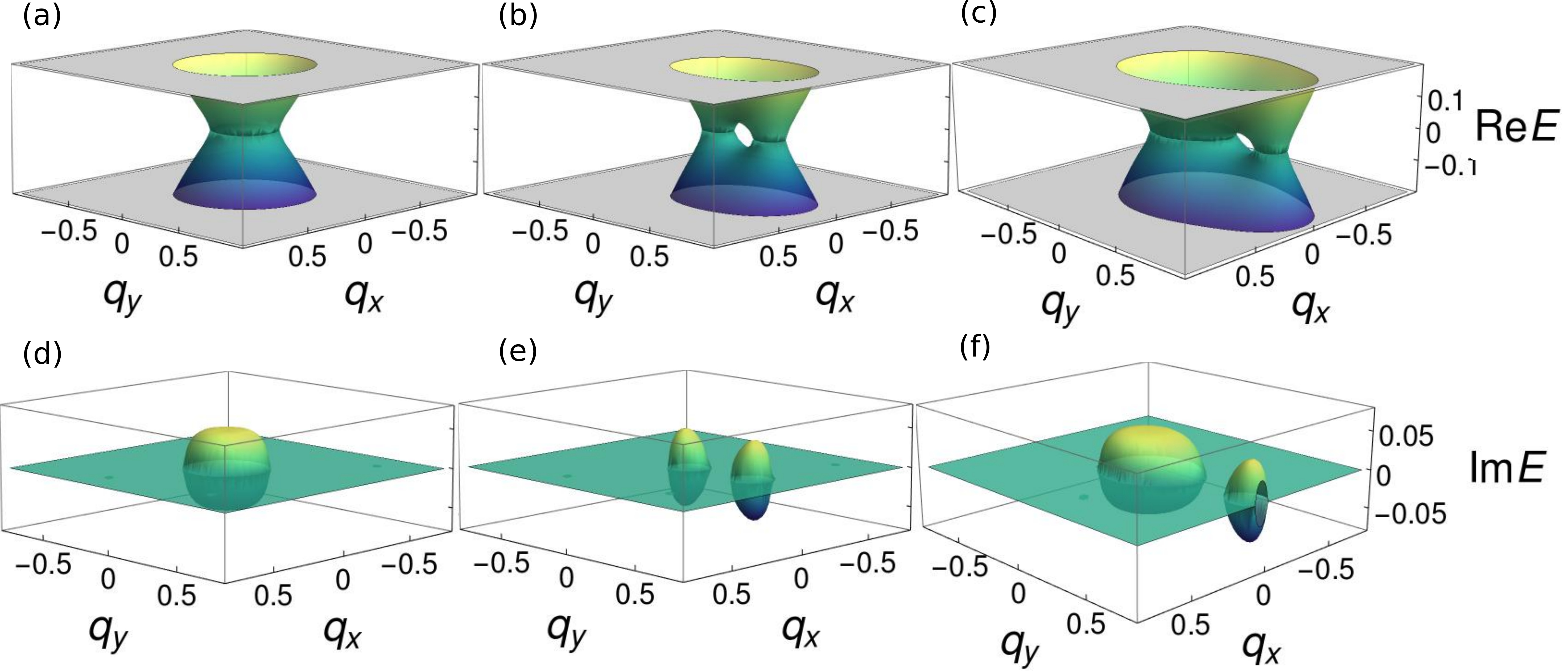}
		\caption{\textbf{Band diagrams showing Lifshitz transitions for multi-WSM.} The real and imaginary parts of the dispersion for double WSM ($n=2$) [(a) and (d)] in the absence of light, and [(b) and (e)] with $A_0=0.77$ and $\zeta=0.08$.  [(c) and (f)] The real and imaginary parts of the dispersion for triple WSM ($n=3$) in the presence of light with $A_0=0.7$ and $\zeta=0.04$. In the absence of light one has a single exceptional ring (ER) for $n=2$ and $n=3$. With increasing light intensity the single ring turns into two ERs resulting in Lifshitz transitions. Here we set  $\Omega=1.0$, $q_z=0.0$, $\eta=1.0$ and $v_z=1.0$.}
		\label{fig2}
	\end{figure}
	
	The band diagrams with topological charge $n=2$ and $n=3$ are shown in Fig.~\ref{fig2}. 
	In general, from Eq.~(\ref{13}), we find the $n$th order polynomial, $f(q)$, governing the locations of the EPs as
	
	\begin{align}\label{f}
	f(q)=  q_{EP_n}^{2n} - 2\Delta_{} q_{EP_n}^{2n - 1} + \Delta_{}^2 q_{EP_n}^{2n - 2} - m^2 \zeta^2=0.
	\end{align}
	
	Using Descartes's sign rule, we find a maximum of three positive real roots and one negative real root. All other roots are imaginary. Thus by tuning the relative strength of the light intensity and the gain and loss parameter, the positions of these four roots can be manipulated. Two or three EPs can be superimposed by adjusting the two parameters. Notably, the condition to achieve the superimposition of $l$ number of EPs is $f^{i}(q_{EP})=0$ for $i=0,1,...,l$. Thus, for $n=2,$ one observes that the two EPs sit together for $\Delta_{}=2\sqrt{\zeta}$. 
	The general criterion for finding the ECs in NH systems is ${\cal E}^{}_{m,+}={\cal E}^{}_{m,-}$, which is equivalent to
	
	\begin{equation}\label{reim}
	\begin{split}
	\quad {\rm Re} \det [H^{\eta,n}(q)] & =0,\\
	\quad {\rm  Im} \det [H^{\eta,n}(q)] & =0.
	\end{split}
	\end{equation}
	
	These two constraint equations (in three-dimensional systems) allow higher-dimensional surfaces to form but are restricted to exhibit only one-dimensional ECs~\cite{Cerjan1}. Along these one-dimensional contours, the eigenvalues coalesce and the eigenvectors also merge. 
	
	\begin{figure}
		\includegraphics[scale=0.1]{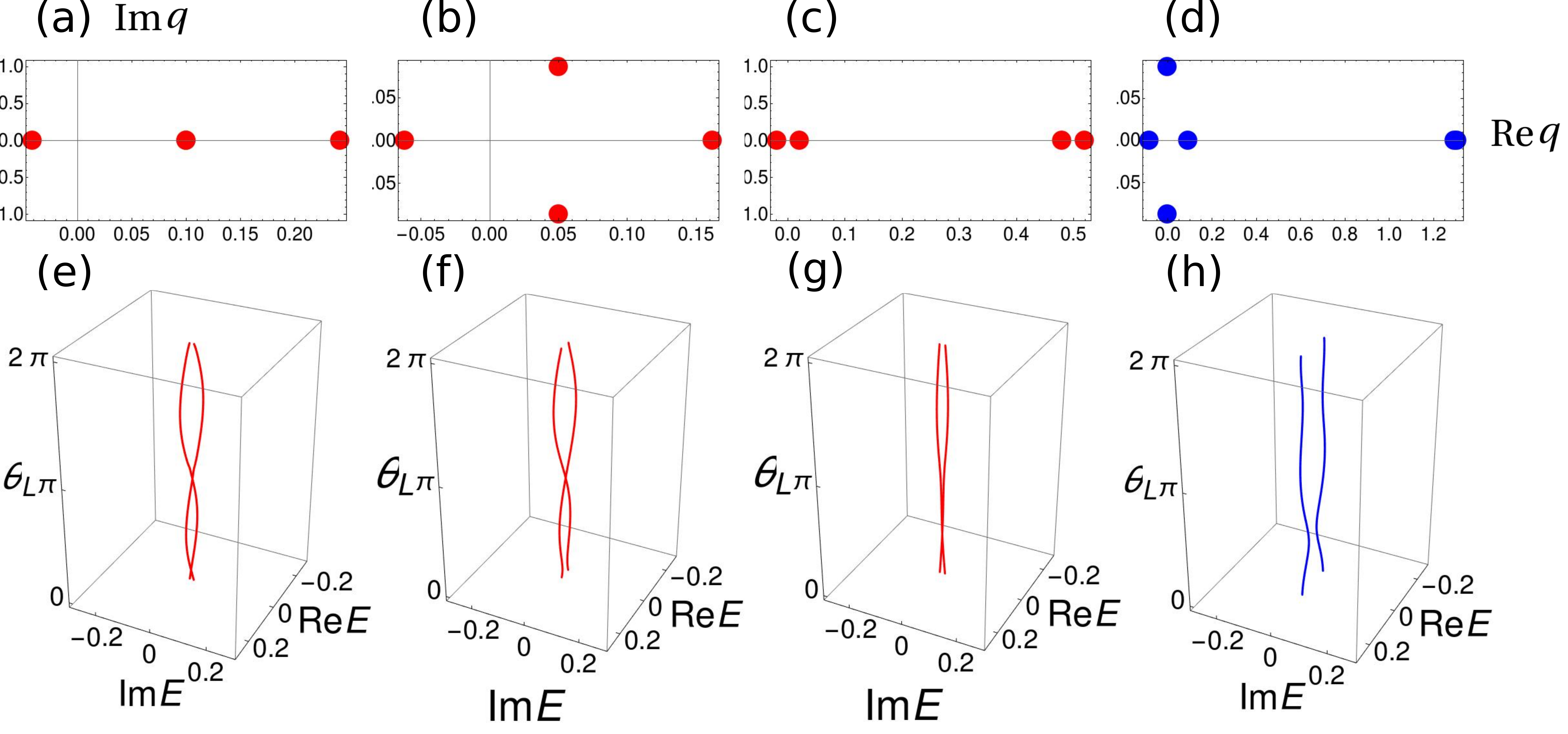}
		\caption{\textbf{Locations of exceptional points and analysis of vorticity.} The locations of [(a), (b) and (c)] four exceptional points for $n=2$ and (d) eight exceptional points for $n=3$ for different light intensities. Corresponding vorticity plots are shown in the lower panel. (e) The vorticity of two bands encircling two exceptional points sitting together with same winding number for $A_0=0.447$ [corresponding to (a)]. Vorticity plots for (f) $A_0=0.32$ and (g) $A_0=0.7$ encircling the left most EP and third EP corresponding to (b) and (c), respectively. (h) The vorticity plot for $n=3$ with $A_0=0.9$ encircling two EPs with the same position and opposite winding numbers showing the absence of swapping. Here we set $\zeta=0.01$.} 
		\label{fig3}
	\end{figure}

	Let us now discuss the specific cases of double ($n=2$) and triple ($n=3$) WSMs in more detail. In case of the double WSM the Floquet effective Hamiltonian can be written as $H_{\mathrm{eff}}^{\eta}(q)=\mathbf{d}(q)\cdot \boldsymbol{\sigma}$, where $\boldsymbol{d}=\mathbf{d}_R +i \mathbf{d}_I$ with $\mathbf{d}_R,\mathbf{d}_I \in \mathfrak{R}^3$ and $\boldsymbol{\sigma}$ the vector of standard Pauli matrices. Using Eq. (\ref{reim}) one obtains the locations of EPs  as follows

	\begin{equation}\label{ep coordinate with light}
	q_{EP}  =\Delta/2\pm\sqrt{\Delta^2/4\pm \zeta},
	\end{equation}
	where we have chosen $\phi=\pi/2$.
	\begin{figure}
		\includegraphics[scale=0.2]{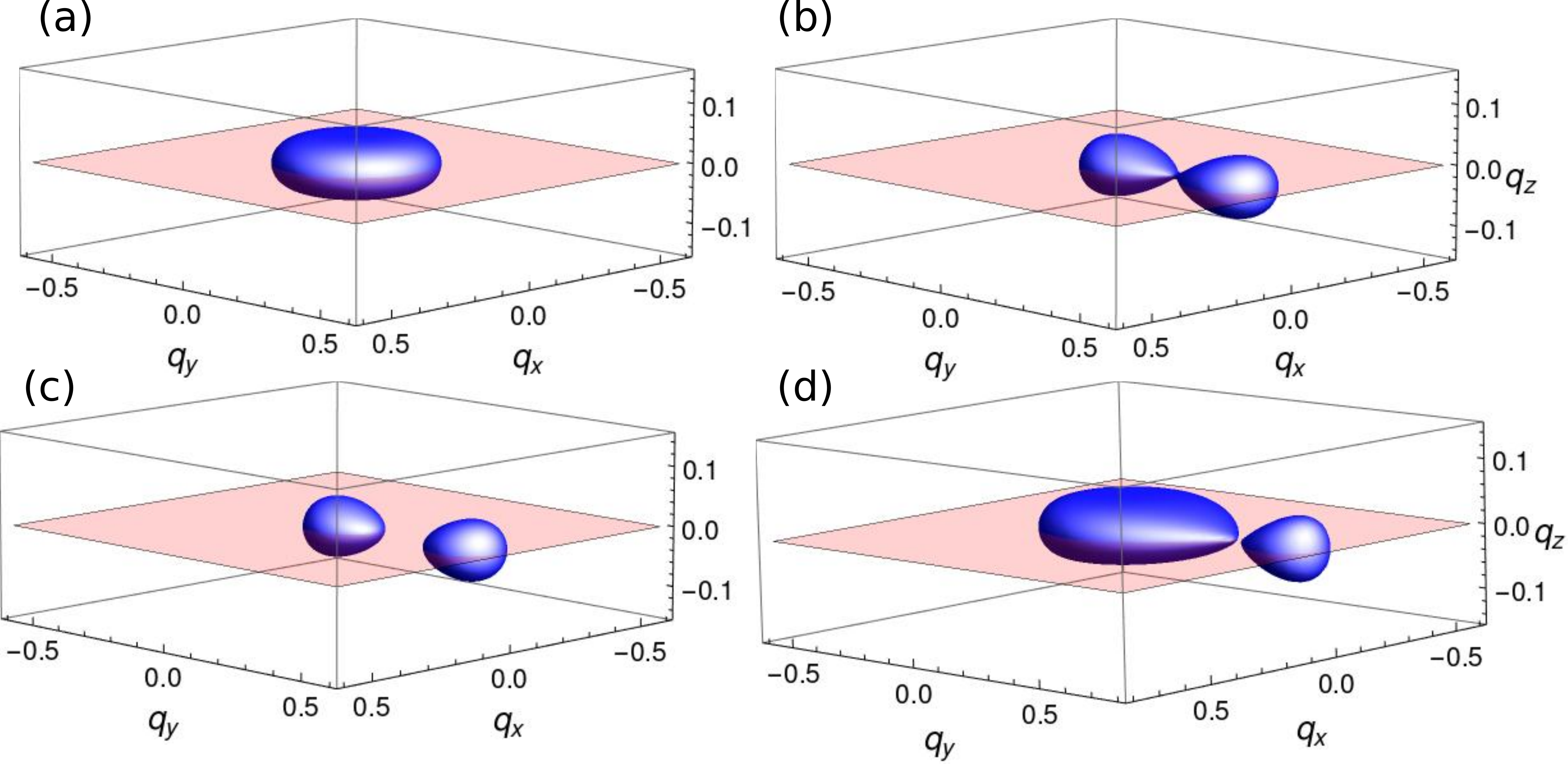}
		\caption{\textbf{Light-tunable exceptional contours for non-Hermitian multi-Weyl semimetals.} The real (blue) and imaginary (orange) exceptional surfaces are shown, and their intersection defines the ECs. ECs for double WSMs ($n=2$) (a) in the absence of light, (b) at critical light intensity with $A_0=0.66$ and (c) at a higher light intensity $A_0=0.68$. (d) EC for $n=3$ is shown for light amplitude $A_0=0.68$. We clearly see the evolution of contours with increasing light intensity, showing topological charge division arising from Lifshitz transitions. Here we set $\zeta=0.05$.}
		\label{fig4}
	\end{figure}
	
	As a consequence of the quadratic dependence of momentum in Eq.~(\ref{13}) for $n=2$, in general we get four EPs in the $q_z=0$ plane. The position of the EPs can be further tuned by varying the light intensity as we present in Fig.~\ref{fig3}. Depending upon the strength of the light field and the NH term, one obtains two EPs (when the momentum coordinates of the other two EPs becomes imaginary), three EPs (when two EPs coincide with each other for a particular value of light intensity, i.e., $\Delta=2\sqrt{\zeta}$).
	
	We present the fate of ECs in the presence of light in Fig.~\ref{fig4} and interestingly, we find that light accomplishes topological charge division by splitting the parent contour symmetrically or asymmetrically, as we will discuss next. For double WSMs, one observes that the single contour is divided into two symmetric contours with the increase of the driving amplitude [Fig.~\ref{fig4}(a)-(c)]. In Fig.~\ref{fig4}(d), we show a similar analysis for a triple WSM. Unlike the double WSM case, where the parent contour is divided into two symmetric contours, in triple WSMs the division is asymmetric. This is due to the fact that driving divides the ECs into those corresponding to a double Weyl and a linear Weyl ones.
	
	An important question next arises: how does the value of the topological charge in the newly generated ECs depend on the driving field? To answer this question, we consider the non-Hermitian generalization of the Berry charge~\cite{Hirsbrunner} as $
	{\mathfrak N}=\int_{C} {\bf\Omega}^{LR}({\textbf k})\cdot d{\bf S},$
	where ${\bf\Omega}^{LR}({\textbf k})$ is the Berry curvature which can be obtained as $\nabla \times  {\cal A}^{LR}({\textbf k}),$ with ${\cal A}^{LR}({\textbf k})$ being the Berry gauge field (see supplementary material for details). ${\cal A}^{LR}$ is obtained from the left and right eigenvectors ($\psi^{L/R}$) of the Hamiltonian as
	${\cal A}^{LR}(k)=i\left<\psi^{L}(k)|\nabla|\psi^{R}(k)\right>.$
	
	\begin{figure}
		\includegraphics[scale=0.3]{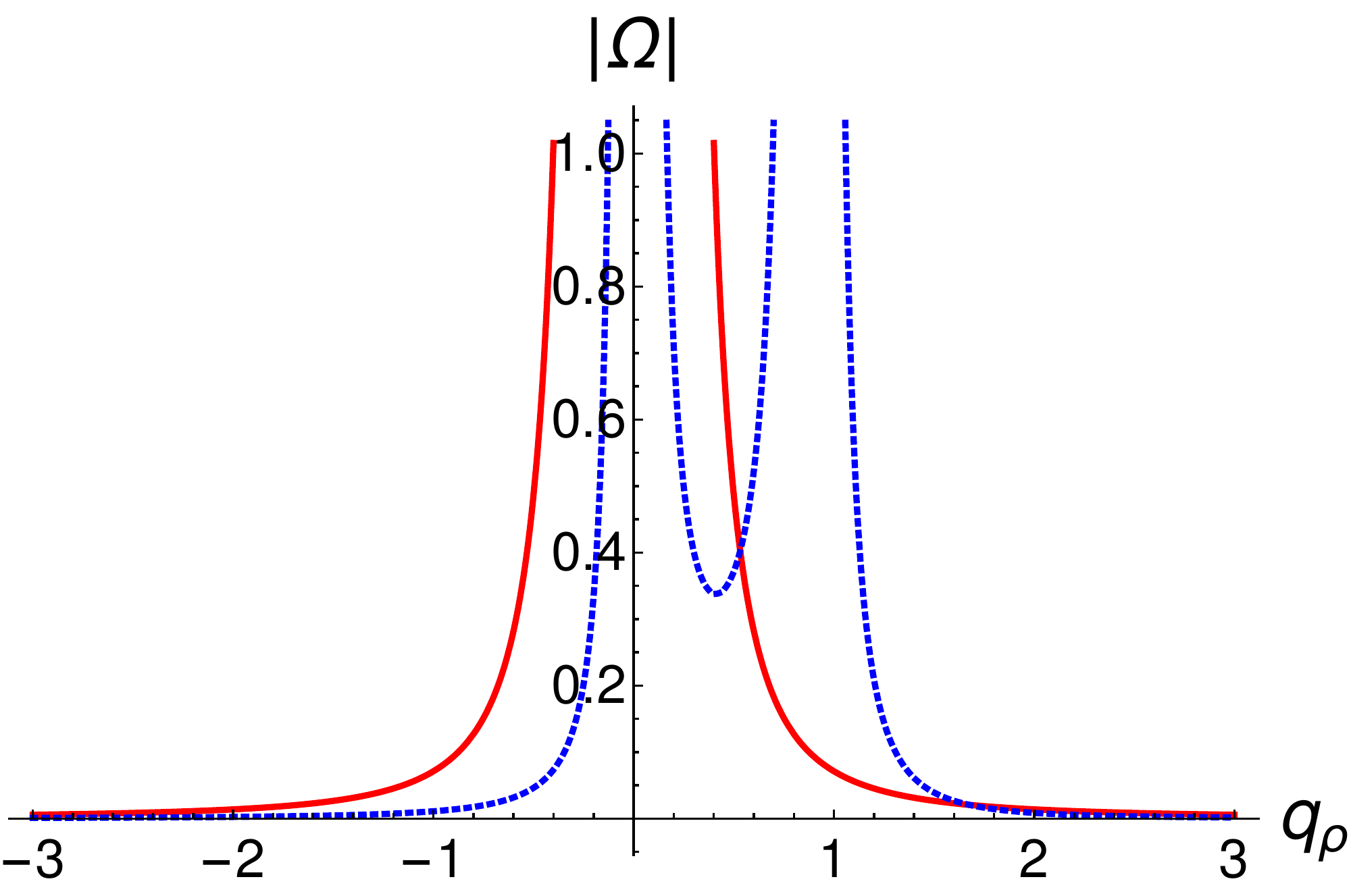}
		\caption{\textbf{Berry curvature density tuning with light.} Normalized Berry curvature density as a function of the radial momentum without light (red solid) and with $A_0=0.8$ and $\zeta=0.05$ (blue dashed). The divergences signal the presence of band degeneracy where the topological charge accumulates. In the absence of light we obtain a single peak, while beyond the critical light intensity the two peaks confirm the charge division. Here $q_x=q_{\rho}\cos{\phi}$ and $q_y=q_{\rho}\sin{\phi}$. We further choose $\phi=\pi/2$.} 
		\label{fig5}
	\end{figure}
	
	We note that it is shown in Ref.~\cite{Cerjan} that after integrating the Berry curvature on a closed surface, which encloses the ECs, the Berry charge is real and quantized. Here we have shown that in presence of the driving field, the charge solely depends on the driving amplitude and when the amplitude crosses a limiting value the charge division occurs. In Fig.~\ref{fig5}, the plot of Berry curvature density for the double WSM is presented. One observes that when light amplitude is less than $0.8$, we obtain only one divergence in the Berry curvature density distribution -- this is the region where the topological charge accumulates. Increase of the driving amplitude causes the ${\bf\Omega}^{LR}({\textbf k})$ to show discontinuity at two different places, which in-turn denotes the accumulation of charges at two different places and results in the division the ECs. Similar argument can be drawn for multi-WSM case as well (see details in supplementary material).
	
	We map out the topological phase diagram next, by computing the vorticity of constitutive bands. One can define the vorticity, $\nu_{mn}$, for any pair of bands ($E_m$ and $E_n$) with complex energy dispersion as~\cite{Shen}
	\begin{equation}\label{vorticity}
	\nu_{mn}(\Gamma)=-\dfrac{1}{2\pi}{\oint_{\Gamma}{{\nabla_{\textbf k}} {\arg[E_m(\textbf k)-E_n(\textbf k)]}} \,d\textbf{k}},
	\end{equation}
	
	where $\Gamma$ is a closed loop encircling the EP in the momentum space. The energy eigenvalue for a single complex band of NH Hamiltonian in general can be written as $E(\textbf k)=\mathopen|E(\textbf k)\mathclose|e^{i\theta_L(k)}$, where $\theta_L=\tan^{-1}({\mathrm{Im}E/\mathrm{Re}E})$. Fractional vorticity is an inherent property of the EPs and is well defined in the absence of any symmetry~\cite{Shen}. We present the vorticity and its tuning by light for our WSM systems in Fig.~\ref{fig3}(e)-(h). When the chosen contour encircles an odd number of EPs, the two bands swap with each other in the complex plane owing to the square root singularity, and the vorticity takes a half-integer value. On the other hand, when an even number of EPs are enclosed, the vorticity becomes an integer~\cite{banerjee2}.

	A complementary diagnostic of the NH topological bands is the winding number, $W_N$. By considering $q_{x},$ $q_{y}$ and $q_{z}$ as parameters one can define it as~\cite{Gong,Yin}
	
	\begin{align}\label{WN}
	W_{N}=\frac{1}{2\pi}\int_{-\infty}^{\infty} dk_{z} \partial{_{k_{z}}}\phi_{xz},
	\end{align}
	
	where $\phi_{xz}=\arctan\left(\frac{h_{x}}{h_{z}}\right)$. Here $h_{x}$ and $h_{z}$ are the components of the Hamiltonian and can be found by comparing Eq. (\ref{12})(for n=2) with $H=h_{x}\sigma_{x}+h_{z}\sigma_{z}$. We find the general expressions for the winding number for multi-WSMs. Depending on whether $n$ is even or odd, different values of the winding number are obtained. The winding number in the $k_{x} =0$ plane is obtained as (see supplementary material for details)
	
	\begin{align}
	W_{N}=\frac{\mathrm{Sgn}\left({\cal B}^{n}+m \zeta\right)+\mathrm{Sgn}\left({\cal B}^{n}-m \zeta\right)}{4},
	\end{align}
	
	where 
	
	\begin{align}
	{\cal B}^{n}=-\Delta k_{y}^{n-1}+k_{y}^{n},~~n=1,2,3....    
	\end{align}
	
	We find a clear dependence of the winding number on the light-induced term. For $n=1,3,5...$ the winding number has values as $\pm 1/2$ for contours enclosing EPs and zero otherwise. On the contrary, for $n=2,4,6...$ the values are $1/2$ for contours which enclose the EP and $0$ otherwise. This further confirms the control over EPs with driving.

	Finally, let us discuss the experimental feasibility of our work. During the last few years, experimental efforts to investigate the role of NH loss/gain on different topological properties in systems such as dissipative wave-guides and cold atom platforms~\cite{Cerjan,Midya,Dembowski,Xu,Luis1} being initiated, where the generation of EPs has been analyzed~\cite{Dembowski,Ding}. In Ref.~\cite{Cerjan}, the creation of ECs was experimentally demonstrated in helical wave-guides, where the existence of Weyl points was observed. Inclusion of some cuts in the wave-guides gives rise to the NH term, which in turn produces the ECs and their real and quantized topological charge has also been measured~\cite{Cerjan}. To observe these ECs, in \cite{Cerjan1}, the authors use metallic chiral woodpile photonic crystals, where Weyl points with topological charges 1 and 2 can be found by introducing complex onsite energy~\cite{Woodpole}. On the other hand, in cold atom systems\cite{CA} the NH and driving aspects are incorporated through atomic population control~\cite{Li} and shaking of an optical lattice~\cite{Jotzu}. Another important direction to experimentally realize our results is to consider a topo-electrical circuit~\cite{NH circuit,NH circuit1,NH circuit2} for multi-WSMs. We have presented a circuit diagram of such a circuit for realization of double-WSMs (see supplementary section). In this set up it is possible to realize the circuit parameters in terms of the Hamiltonian of the double WSMs. The experimental detection of nodal band structures is possible by capturing the complex admittance spectra for each fixed $q_y$ and striking changes can be observed in complex admittance spectra at ${q_{EP}}$, which enables the detection of the ERs and changes in the Fermi surface. Here the Floquet term could easily be incorporated by switching on and off the circuit abruptly. It is important to note that the circuit elements can be reliably time-modulated
		through the use of MOSFETs that allow component parameters, i.e., inductance to be switched via external control voltages. 
	
	In summary, we have explored the role of driving with light in NH multi-WSMs. We have illustrated our proposal of how driving allows control over existing ECs, as well as enables spawning new ECs. This is one of the main highlights of our work. We have further pointed out that light allows tuning of distribution of topological charge as well as Lifshitz transitions. The charge division is demonstrated through the analysis of the NH generalization of the Berry curvature, which shows a single discontinuity when the light amplitude is less than a critical value. Upon increasing the light amplitude, one comes across a discontinuity of the Berry curvature at two different values of the momentum -- this signals a splitting of the EC and a division of the topological charge. Furthermore, we have diagnosed the exceptional physics by means of vorticity and winding number computations. We hope our results motivate future theoretical and experimental investigation of interplay between driving and non-Hermiticity.

	\noindent \textit{Acknowledgments:} D.C. acknowledges financial support from DST
	(project number SR/WOS-A/PM-52/2019). A.B. thanks Indian Institute of Science for a fellowship. A.N. acknowledges support from a startup grant (SG/MHRD-19-0001) of the Indian Institute of Science and DST-SERB (project number SRG/2020/000153).

	\appendix
	\newpage
	\begin{widetext}
		
		\textbf{Supplementary material: Light-driven Lifshitz transitions in non-Hermitian multi-Weyl semimetals}\\
		
		\section{Floquet-Magnus expansion for driven non-Hermitian multi-Weyl semimetals}
		In this section we would present the computational details of the high frequency approximated effective Hamiltonian. 
			We are dealing with a situation where the system with loss/gain is driven by a circularly polarized light. The Hamiltonian of a multi-Weyl semimetal in presence of a loss/gain term is
			\begin{equation}
			\label{A1}
			H^{\eta,n}(q)=\frac{1}{2m}\Big(q_{-}^{n}\sigma_{+}+q_{+}^{n}\sigma_{-}\Big)+\left(\eta v_z q_z+i\zeta\right)\sigma _z,
			\end{equation}
			where $q_{\pm}=q_{x}\pm i q_{y}$, $\sigma_{\pm}=\sigma_{x}\pm i \sigma_{y}$, $\eta=\pm 1$ and $\zeta$ is the gain/loss parameter.
			Now we are interested in a situation where the multi-Weyl semimetal is driven by a circularly polarized light polarized in the $xz$ plane. In that case the momentum in the Hamiltonian (\ref{A1}) changes as
			\begin{align}
			q_x\rightarrow q_x+A_{0} \sin(\Omega t),\nonumber\\
			q_z\rightarrow q_z+A_{0} \cos(\Omega t),	
			\end{align}
			where $A_{0}$ and $\Omega$ are the driving field amplitude and frequency respectively.
			In this case we can write the resulting Hamiltonian as follows
			\begin{align}
			H_{1}(t)=H^{\eta,n}(q)+{\cal V}(t),
			\end{align}	
			where ${\cal V}(t)$ is the time dependent part of the Hamiltonian. In this case
			\begin{align}
			{\cal V}(t)=\frac{n A_{0}}{2m} \sin(\Omega t)\Big(q_{-}^{n-1}\sigma_{+}+q_{+}^{n-1}\sigma_{-}\Big)+\eta v_{z}A_{0} \cos(\Omega t).
			\end{align}
			It is important to introduce the high frequency expansion here to discuss the effect of driving. In such an expansion, we search for an effective Hamiltonian which describes the dynamics of the system on a time scale which is much longer than the time period $T$. In this scenario, the response can be described by an average over a period. Following \cite{He}, one can simply write the effective Floquet Hamiltonian as time independent and the structure of the effective Hamiltonian upto order $1/\Omega$ is
			\begin{align}\label{A5}
			H_{\mathrm{eff}}=H^{\eta,n}(q)+\frac{1}{\hbar \Omega}\Big[{\cal V}_{-1},{\cal V}_{1}\Big],
			\end{align} 
			where 
			\begin{align}\label{A6}
			{\cal V}_{\pm 1}=\frac{1}{T}\int_{0}^{ T}{\cal V}(t)e^{\pm i\Omega t},
			\end{align}
			with $T=\frac{2\pi}{\Omega}$. One should note here that the effective Hamiltonian we get is exactly similar to the hermitian case. Following \cite{He}, one can argue that upto $1/\Omega$ term, the expansion gives similar terms in the effective Hamiltonian as that of the hermitian case.
			Using Eqs. (\ref{A5}) and (\ref{A6}), we can write
			\begin{align}\label{A7}
			H_{\mathrm{eff}}^{n,\eta}(q)=\frac{1}{2m}\Big((q_{-}^{n}+i \Delta_{}q_{-}^{n-1}) \sigma_{+}+(q_{+}^{n}-i \Delta_{}q_{+}^{n-1})\sigma_{-}\Big)+\left(\eta v_{z} q_{z}+i\zeta\right)\sigma _z,
			\end{align}
			where $\Delta_{} =\frac{nA_{0}^{2}v_{z}\eta}{2\hbar \Omega}.$
			This is Eq. (2) of the main text.		
		\section{Note on the symmetry of multi-Weyl semimetals}

		\begin{figure}\label{figa1}
			\includegraphics[scale=0.8]{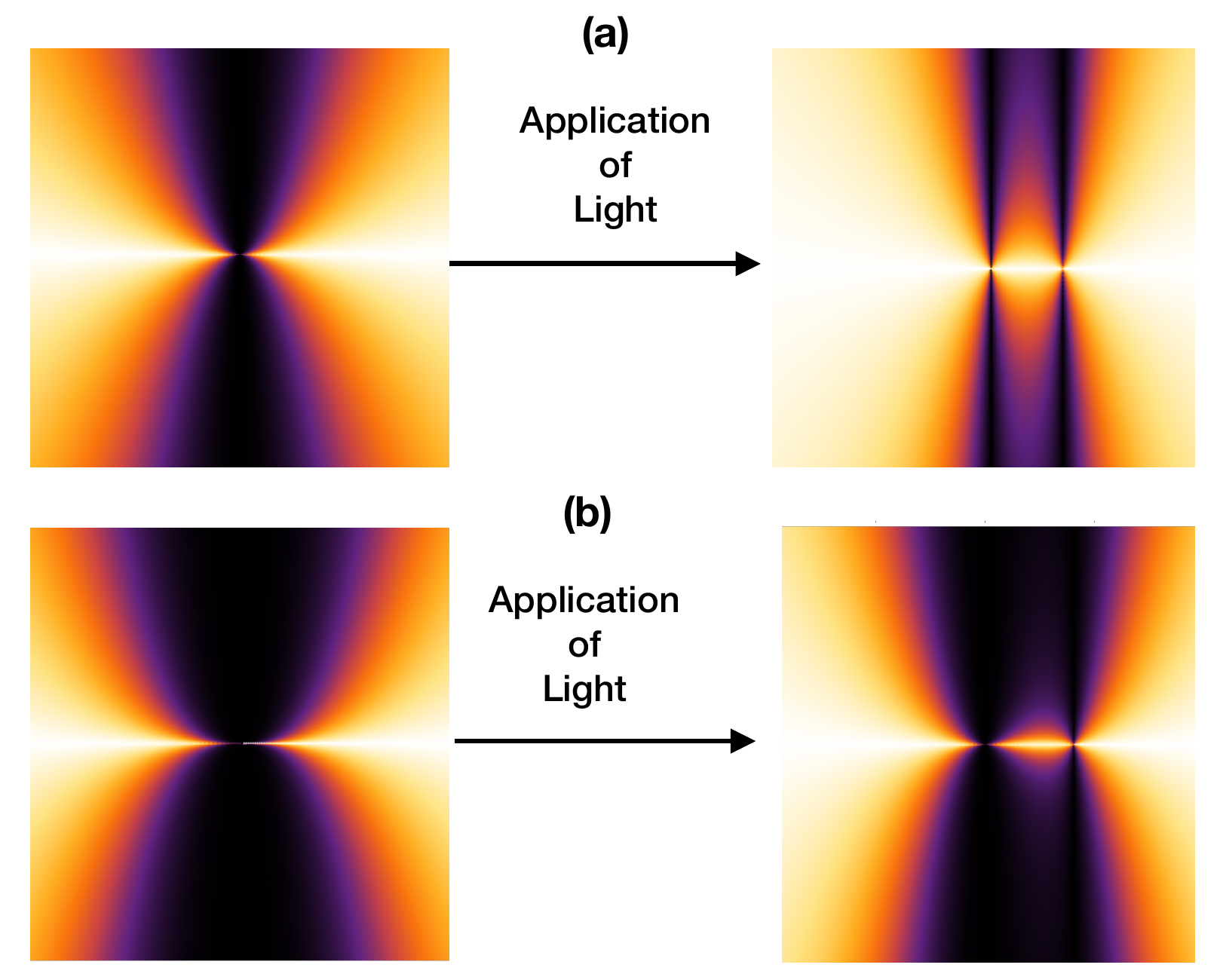}
			\caption{\textbf{Effect of light on multi Weyl semimetals.} (a) The division of the charge 2 double Weyl semimetal into two charge 1 Weyl points in absence of any loss/gain term and (b) The division of the triple Weyl semimetal into one double WSM and one single Weyl point in absence of any loss/gain term. The vertical axis is the energy and the horizontal axis is the momentum.}
		\end{figure}

		In this section, we would like to make further comments on multi-Weyl semimetals (multi-WSMs). Let us first begin with the case of unperturbed double WSM. In this case as we have mentioned in the main text, one finds a non-linear, parabolic dispersion. This kind of WSMs are found experimentally in systems such as strontium silicide (SrSi$_{2}$)~\cite{MW2}. It was shown previously that the strong spin orbit coupling (SOC) of this material is responsible for this unusual feature. It is observed that without SOC, SrSi$_{2}$ shows features of a usual charge $\pm 1$ WSM. Inclusion of the SOC term binds two linearly dispersive Weyl Fermions of same charge together and form a double WSM with topological charge $\pm 2$. This kind of double WSM is protected by point groups $C_{4}$ or $C_{6}$~\cite{MW1,MW2}. In case of a triple WSM, the corresponding rotational symmetry is $C_{6}$~\cite{TW}. As the multi-Weyl nodes considered in our work are protected by $C_{n}$, an applied strain induces a phase transition by breaking the rotational symmetry. This results in splitting of multi WSMs into several single WSMs with topological charge $\pm 1.$ In Ref.~\cite{MW1,MW2}, the authors have shown the splitting of such a multi WSM in presence of an anisotropic strain. 
		The triple WSMs also split into three charge $\pm 1$ WSMs, in presence of $C_{6}$ rotational symmetry breaking terms.
		
		In this paper we have shown that driving with a circularly polarized light causes the division of a double WSM into two single WSMs. The division of a double WSM (triple WSM) into its constituents is shown in Fig. 1. In addition to the driving field, when one adds the gain/loss term, which is the main focus of our work, the Weyl points form exceptional rings and as a result, new exceptional contours are generated. This generation of controllable exceptional contours is one of the main results of our paper.
		
		Another unique result of the paper is the charge distribution among the newly generated exceptional contours by application of the driving. This analysis demands a complete discussion of Berry curvature, which we have presented next in Sec. II.
		
		
		\section{Analysis of the Berry Curvature}
		
		In this section, we would like to include the detailed analysis of the Berry curvature. The left and right eigen-vectors of the multi-Weyl semimetal are written as
		
		\begin{align}\label{B1}
		\Big<\psi^{L}\Big|&= \left(\frac{\Delta_{m}  q_{\rho}^{n-1} \sin ((n-1) \phi )+\Delta_{m}  q_{\rho}^{n-1} \cos ((n-1) \phi )-i q_{\rho}^n \sin (n \phi )+q_{\rho}^n \cos (n \phi )}{ {\mathfrak L}_{p}},\frac{\lambda -(q_{z}+i \zeta )}{ {\mathfrak L}_{p}}\right),\nonumber\\
		\Big|\psi^{R} \Big>&= \left(\frac{\Delta_{m}  q_{\rho}^{n-1} \sin ((n-1) \phi )+\Delta_{m}  q_{\rho}^{n-1} \cos ((n-1) \phi )+i q_{\rho}^n \sin (n \phi )-q_{\rho}^n \cos (n \phi )}{ {\mathfrak L}_{p}},\frac{\lambda -(q_{z}+i \zeta )}{ {\mathfrak L}_{p}}\right)^{T},
		\end{align}
		
		where we have considered the cylindrical coordinates ($q_{\rho},\phi,q_{z}$) and have also incorporated $q_{\pm} =q_{\rho} e^{\pm i \phi}$. In Eq. (\ref{B1}) we denote
		
		\begin{align}\label{B2}
		\lambda&= \frac{ \sqrt{\Big(\mathfrak{G}_{n}(\phi)+ q_{\rho}^{2 n+4}-\zeta ^2 q_{\rho}^4+ q_{\rho}^4 q_{z}^2 +2 i \zeta  q_{\rho}^4 q_{z} }\Big)}{{q_{\rho}^2}},\nonumber\\
		{\mathfrak L}_{p}&= \sqrt{2 \lambda ^2-2 \lambda  (q_{z}+i \zeta )},
		\end{align}
		
		where
		
		\begin{align}
		\mathfrak{G}_{n}(\phi)= -\Delta_{m} ^2  q_{\rho}^{2 n+2} \cos (2 (n-1) \phi )+(1+i) \Delta_{m} q_{\rho}^{2 n+3} \sin ((2n-1) \phi )-(1-i) \Delta_{m}  q_{\rho}^{2 n+3} \sin \phi.
		\end{align}
		
		Now on we discuss the specific case of $n=2$ and for this case we obtain
		
		\begin{align}\label{B3}
		\lambda= &\frac{1}{{q_{\rho}^2}}\Big(\sqrt{\mathfrak{G}_{2}(\phi) + q_{\rho}^{8}-\zeta ^2 q_{\rho}^4+ q_{\rho}^4 q_{z}^2 +2 i \zeta  q_{\rho}^4 q_{z} }\Big),
		\end{align}
		
		where $\mathfrak{G}_{2}(\phi)$ is the value of the function for $n=2.$
		The Berry gauge field is calculated as ${\cal A}^{LR}(k)=i\left<\psi^{L}(k)|\nabla|\psi^{R}(k)\right>$ and using which we calculate the Berry curvature. For computational purposes we consider $\phi=\pi/2$ and obtain
		
		\begin{align}\label{B4}
		\Omega_{q_{\rho}} &=-\frac{q_{\rho}^5 (q_{\rho}-\Delta ) (2 q_{\rho}+i \Delta ) \left(q_{\rho}^6-2 \Delta  q_{\rho}^5+\Delta ^2 q_{\rho}^4+2 q_{\rho}^2 (q_{z}+i \zeta )^2-2 (q_{z}+i \zeta ) {\mathfrak D}\right)}{2 {\mathfrak D} \left(q_{\rho}^6-2 \Delta  q_{\rho}^5+\Delta ^2 q_{\rho}^4+q_{\rho}^2 (q_{z}+i \zeta )^2-(q_{z}+i \zeta ) {\mathfrak D}\right)^2},\nonumber\\
		\Omega_{\phi} &=0,\nonumber\\
		\Omega_{q_{z}}&=\Big[q_{\rho}^4 \Big(q_{\rho}^6 \left((2+i) \Delta  {\mathfrak D}-2 {\mathfrak D}-(6+(2+i) \Delta ) \Delta ^2 (q_{z}+i \zeta )\right)+2 \Delta  q_{\rho}^5 \left(-(2+i) \Delta  {\mathfrak D}+(2-i) {\mathfrak D}+\Delta ^2 (q_{z}+i \zeta )\right)\nonumber\\&+q_{\rho}^4 \left((2+i) \Delta^3 {\mathfrak D}+(-2+5 i) \Delta ^2 {\mathfrak D}-(6+(2+i) \Delta ) (q_{z}+i \zeta )^3\right)+2 \Delta  q_{\rho}^3 \left((2-i) (q_{z}+i \zeta )^3-2 i \Delta ^2 {\mathfrak D}\right)\nonumber\\&+q_{\rho}^2 \left(i \Delta ^4 {\mathfrak D}-(2+i) \Delta  {\mathfrak D} \zeta^{2}-6 {\mathfrak D} \zeta ^2+(2+i) \Delta  {\mathfrak D} q_{z}^{2}+6 {\mathfrak D} q_{z}^2-(2-4 i) \Delta  {\mathfrak D} \zeta  q_{z}+12 i {\mathfrak D} \zeta  q_{z}+i \Delta ^2 (q_{z}+i \zeta )^3\right)\nonumber\\&-(4-2 i) \Delta  {\mathfrak D} q_{\rho} (q_{z}+i \zeta )^2-i \Delta ^2 {\mathfrak D} (q_{z}+i \zeta )^2-(2+(2+i) \Delta ) q_{\rho}^8 (q_{z}+i \zeta )\nonumber\\&+2 \Delta  (3+(2+i) \Delta ) q_{\rho}^7 (q_{z}+i \zeta )\Big)\Big]/\Big(2 {\mathfrak D} \left(-{\mathfrak D} (q_{z}+i \zeta )+q_{\rho}^6-2 \Delta  q_{\rho}^5+\Delta ^2 q_{\rho}^4+q_{\rho}^2 (q_{z}+i \zeta )^2\right)^2\Big),
		\end{align}
		
		where
		
		\begin{align}\label{B5}
		{\mathfrak D}= \sqrt{q_{\rho}^4 \left(q_{\rho}^2 (q_{\rho}-\Delta )^2+(q_{z}+i \zeta )^2\right)}.
		\end{align}
		
		Here $\Omega_{q_{\rho}},$ $\Omega_{\phi}$ and $\Omega_{q_{z}}$ denote the components of the Berry curvature along $q_{\rho},~\phi$ and $q_{z}$ directions respectively. 
		From the components of the Berry curvature we have calculated the absolute value as follows,
		
		\begin{align}\label{B6}
		|\Omega|= \sqrt{|\Omega_{q_{\rho}}|^{2}+ |\Omega_{\phi}|^{2}+|\Omega_{q_{z}}|^2}.
		\end{align}
		
		Using Eq. (\ref{B6}), we have plotted Fig. (4) of the main text. A complementary approach to understand the topological nature of these non-Hermitian systems is through the winding number, which we discuss in the next section. 
		
		\section{Winding number for double Weyl semimetals}
		
		Let us now calculate the winding number for the driven double WSM. Without loss of generality, we consider a path in the $k_{x} =0$ plane. In doing so the effective Hamiltonian in Eq. (7) of the main text can be written as
		
		\begin{align}\label{29}
		H_{\mathrm{eff}}^{\eta}(k)& =\frac{-k_y^2+\Delta_{} k_{y}}{m}\sigma _x+\left(\eta v_zk_z+i\zeta \right)\sigma _z.
		\end{align}
		
		The winding number is
		
		\begin{align}\label{30}
		W_{N}=\frac{1}{2\pi}\int_{-\infty}^{\infty} dk_{z} \partial{_{k_{z}}}\phi_{yz},
		\end{align}
		
		where
		
		\begin{align}
		\phi_{yz}=\arctan\left(\frac{h_{y}}{h_{z}}\right)=-\arctan\left(\frac{k_{y}^{2}-\Delta k_{y}}{m\left(\eta v_zk_z+i\zeta \right)}\right),
		\end{align}
		
		is a complex quantity and is written as $\phi_{yz}=\phi_{R,yz}+\phi_{I,yz}$. We need to know the value of $\phi_{yz}$ at $k_{z}\rightarrow \pm \infty$, which is calculated to be
		
		\begin{align}\label{32}
		\phi_{yz}(k_{z}\rightarrow\pm \infty)=0_{\pm}   
		\end{align}
		
		Following \cite{Banerjee, Yin}, we can write 
		
		\begin{align}
		e^{-2\phi_{I,yz}}=\frac{\left(\eta v_zk_z+i\zeta \right)-i(k_{y}^{2}-\Delta k_{y})}{\left(\eta v_zk_z+i\zeta \right)+i(k_{y}^{2}-\Delta k_{y})}.
		\end{align}
		
		It is evident that $\phi_{I,yz}$ is an odd function of $k_{z}.$ Thus
		
		\begin{align}
		\frac{1}{2\pi}\int_{\infty}^{\infty} dk_{z} \partial{_{k_{z}}}\phi_{I,yz}=0.
		\end{align}
		
		Then, the real part of $\phi_{R,yz}$ can be written as 
		
		\begin{align}\label{35}
		\tan(2\phi_{R,yz})=\tan(\phi_{A,yz}+\phi_{B,yz}),
		\end{align}
		
		where
		
		\begin{align}
		\tan (\phi_{A,yz}) &=\frac{\zeta-\frac{k_{y}^{2}-\Delta k_{y}}{m}}{\eta v_{z}k_{z}},\nonumber\\
		\tan (\phi_{B,yz}) &=-\frac{\zeta+\frac{k_{y}^{2}-\Delta k_{y}}{m}}{\eta v_{z}k_{z}}.
		\end{align}
		
		From Eq. (\ref{35}) we can write the following
		
		\begin{align}
		\phi_{R,yz}=n\pi+\frac{(\phi_{A,yz}+\phi_{B,yz})}{2}.
		\end{align}
		
		This gives,
		
		\begin{align}\label{38}
		\phi_{A,yz}(k_{z}\rightarrow 0_{\pm})&=\pm\frac{\pi}{2}\mathrm{Sgn}\left(\Delta k_{y}-k_{y}^{2}+m \zeta\right),\nonumber\\  
		\phi_{B,yz}(k_{z}\rightarrow 0_{\pm})&=\pm\frac{\pi}{2}\mathrm{Sgn}\left(\Delta k_{y}-k_{y}^{2}-m \zeta\right).
		\end{align}
		
		Eq. (\ref{30}) is rewritten as
		
		\begin{align}
		W_{N} &=\frac{1}{2\pi}\int_{-\infty}^{\infty} dk_{z} \partial{_{k_{z}}}\phi_{R,yz}\nonumber\\
		&=-\frac{1}{4\pi}\Bigg[2\frac{\pi}{2}\mathrm{Sgn}\left(\Delta k_{y}-k_{y}^{2}+m \zeta\right)+2\frac{\pi}{2}\mathrm{Sgn}\left(\Delta k_{y}-k_{y}^{2}-m \zeta\right)\Bigg]\nonumber\\
		&=\frac{\Bigg[\mathrm{Sgn}\left(k_{y}^{2}-\Delta k_{y}-m \zeta\right)+\mathrm{Sgn}\left(k_{y}^{2}-\Delta k_{y}+m \zeta\right)\Bigg]}{4}
		\end{align}
		
		This gives
		
		\begin{align}
		W_{N} &= \frac{1}{2},~~~ k_{y}<\frac{\Delta}{2}\pm \sqrt{\frac{\Delta^{2}}{4}-m\zeta}\nonumber\\
		&=\frac{1}{2},~~~ k_{y}>\frac{\Delta}{2}\pm \sqrt{\frac{\Delta^{2}}{4}+m\zeta}\nonumber\\
		&=0,~~~ \mathrm{otherwise.}
		\end{align}
		
		\section{Experimental realizations of the Nodal exceptional structures and Circuit realization of non-Hermitian double Weyl semimetals}
		
		In this section we would briefly propose the possible experimental techniques to realize the nodal structures. In \cite{Cerjan} the authors have shown a procedure to achieve the exceptional contours (ECs) in photonic crystals. To observe these ECs, the authors use metallic chiral woodpile photonic crystals, where Weyl points with topological charges 1 and 2 can be found by introducing complex onsite energy~\cite{Woodpole}. The merging of contours can be realized by tuning the onsite potential of photonic crystal operated in the terahertz frequency regime.
			Very recently, NH Weyl phase has been proposed in 3D topological insulators coupled to a feromagnetic lead and the non-Hermitian physics of this Weyl phase can be easily tuned with the magnetization direction of the ferromagnetic lead~\cite{FM}. Most importantly, the NH Weyl physics and concomitant nodal band structure, discussed by authors are amenable to surface spectroscopy, robust against perturbation. Thus shinning light on such material junction can be a readily tunable platform for observation and manipulation of light induced NH topological phases~\cite{FM}.
		
		Another way of realizing our formalism could be optically shaken cold atom systems introducing loss through selective depopulation of cold atoms. The required potential for the double Weyl band structure can be achieved by tuning time dependent oscillations controlled interfering laser beams. The tight-binding model comprising of various kinds of intralayer and interlayer hoppings can achieved from the overlap integral between atomic orbitals. At the same time, the minimal momentum coupling can be engineered by directional shaking. The high frequency regime of Magnus expansion can be accomplished by small oscillation amplitudes which, in turn, gives a static Hamiltonian creating sublattices and pseudospins. Thus, our proposed nodal band structures can also be realized in cold atom systems~\cite{CA}.
		
		Besides, the experimental feasibility of our work can be realized by considering a topo-electrical circuit \cite{NH circuit,NH circuit1,NH circuit2}.
			We consider a convenient synthetic platform to realize multi-Weyl semimetal (specially for double Weyl( $n=2$)) based on AC circuits consisting of periodic arrays of capacitors and inductors~\cite{NH circuit,NH circuit1,NH circuit2}. The schematic diagram of the circuit is presented in Fig.~\ref{circuit_diag}.

		The Hamiltonian of the double Weyl semimetal is given as
			\begin{align}
			H_{} =\textbf{d}\cdot {\boldmath\sigma}
			\end{align}
			where
			\begin{align}\label{d}
			d_x &= 2t(1-\cos{q_x})+2t(1-\cos{q_y})=t_{1}'-t_{1}\Big(e^{iq_{x}}+e^{-iq_{x}}\Big)-t_{1}\Big(e^{iq_{y}}+e^{-iq_{y}}\Big),\nonumber\\
			d_y &= v\sin{q_x}\sin{q_y}=-\frac{v}{4}\Big(e^{iq_{x}}-e^{-iq_{x}}\Big)-t_{1}\Big(e^{iq_{y}}-e^{-iq_{y}}\Big),\nonumber\\
			d_z &= i\gamma.   
			\end{align}

		In real space the Eq. (\ref{d}) reads
			\begin{align}
			d_x &= \sum_r   t_{1}^{'} a_r^{\dagger} b_r - t_{1}^{'}a_r^{\dagger}\Big( b_{r+x}+b_{r-x}\Big) +t_{1}^{'}a_r^{\dagger}\Big( b_{r+y}+b_{r-y}\Big)+{\rm h.c.},\nonumber\\
			d_y &=-\frac{v}{4}\sum_r a_r^{\dagger}\Big( -b_{r+x+y}+b_{r+x-y}+b_{r-x+y}-b_{r-x-y}\Big) +{\rm h.c.},\nonumber\\
			d_z &= i\gamma\sum_r\Big(a_r^{\dagger} b_r+b_r^{\dagger} a_r\Big).
			\end{align}
			Here $\mathrm{h.c.}$ denotes the corresponding Hermitian conjugate term and $r$ is the site label in real space.

		The corresponding Lagrangian is
			\begin{align}
			L&=\frac{1}{2}\sum_{r}\Big[ C^0\Big((\dot\varphi_r^A - \dot\varphi_{r-\hat y}^B)^2+(\dot\varphi_r^A - \dot\varphi_{r+\hat y}^B)^2\Big) -\frac{1}{L_r^0}(\varphi_r^A - \varphi_{r}^B)^2 -\frac{1}{L_{r}^1}\Big((\varphi_r^A - \varphi_{r+\hat x}^B)^2+(\varphi_r^A - \varphi_{r-\hat x}^B)^2\Big)\nonumber\\& +C^1\Big(-(\dot\varphi_r^A - \dot\varphi_{r+\hat x+\hat y}^B)^2+(\dot\varphi_r^A - \dot\varphi_{r+\hat x-\hat y}^B)^2-(\dot\varphi_r^A - \dot\varphi_{r-\hat x-\hat y}^B)^2+(\dot\varphi_r^A - \dot\varphi_{r-\hat x+\hat y}^B)^2\Big)\Big],
			\end{align}
		\begin{figure}
			\includegraphics[scale=0.1]{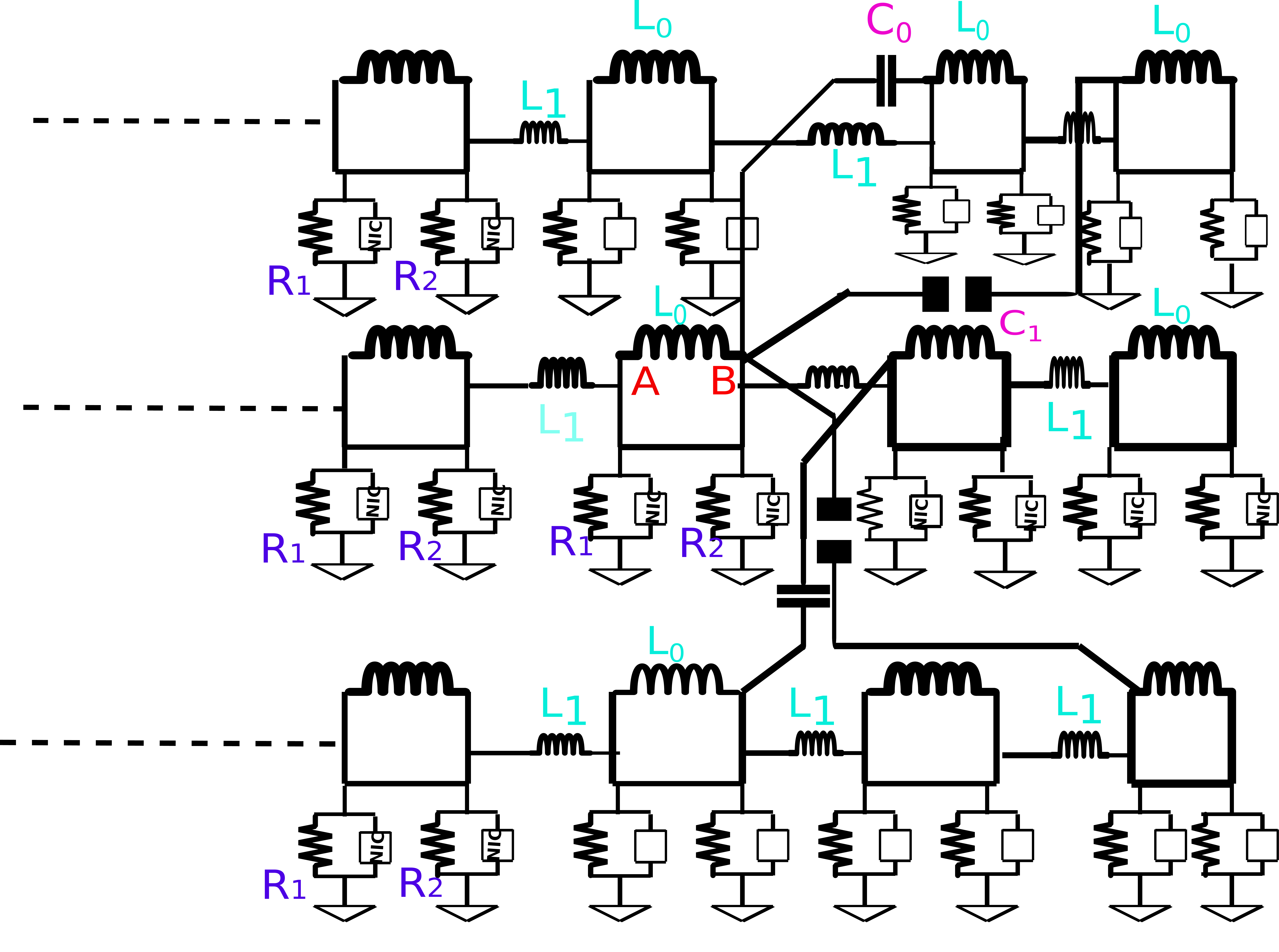}
			\caption{\textbf{Schematic diagram of the topo-electric circuit for a double-Weyl semimetal.}  The nearest neighbour and next-nearest neighbour couplings at right half of the A, B unit cell are shown in the diagram. The dotted lines indicate the presence of a similar structure in the left hand side, which completes the circuit for the double Weyl semimetal.} \label{circuit_diag}
		\end{figure}
		
	with the Rayleigh dissipation function
			\begin{equation}\label{eq:D}
			D=\frac{1}{2}\sum_r\Big(- \frac{1}{R_{1}}(\dot\varphi_r^A)^{2} - \frac{1}{R_{2}}(\dot\varphi_{r}^B)^2 - \frac{1}{R}({\dot\varphi_r^{A\,2}} + {\dot\varphi_r^{B\,2}})\Big).
			\end{equation}
			The second term in $D$ accounts for negative static converter (NIC) resulting in static resistance in the circuit. Based on the Kirchoff current law one can form an admittance problem solving the equation of motion from the Lagragian for a given flux at a particular frequency. The admittance matrix for two atom unit cell (inequivalent nodes $A$ and $B$) dictates the voltage response in the circuit connecting currents at different nodes. The different kinds of intra-layer and inter-layer hoppings can be realized by capacitors and inductors, which resembles the tight binding Hamiltonian for non-Hermitian double Weyl semimetal. We can tune the strength of non-Hermiticity by using unequal resistors grounding the inequivalent nodes.

		Following Ref.~\cite{NH circuit}, we can write different components of the Hamiltonian in terms of the circuit parameters as follows
			\begin{align}\label{alld}
			d_x &=  -\frac{1}{\omega L^0} +2\omega C^0\cos{q_y} - \frac{2}{\omega L^1}\cos{q_x}\,, \qquad
			d_y = 4\omega C^1\sin{q_y}\sin{q_x},\nonumber \\
			d_0 &= -2\omega c_0+ \frac{1}{\omega} \Big[ (\frac{1}{L^0}+\frac{2}{L^1}\Big] +\frac{i}{2} \Big[(\frac{1}{R_A}+\frac{1}{R_B})\Big] \,,\qquad
			d_z = \frac{i}{2}\Big(\frac{1}{R_{A}}-\frac{1}{R_{B}}\Big).
			\end{align}
			Here $d_0$ is the chemical potential term, which only shifts the energy, and can be tuned by the circuit parameters. The eigenvalues of the Hamiltonian are complex in the presence of $\gamma$ (non-zero $R$) and the Weyl points turn into exceptional rings consisting of exceptional points at $q_{EP}$ along $q_z=0$ line, where the corresponding Hamiltonian is non-diagonalizable.
		
		The experimental detection of nodal band structures can be verified at a particular resonance frequency fed by a sine wave generator and by tracing the complex admittance spectra for each fixed $q_y$. One can observe distinct changes in complex admittance spectra exhibiting genus surfaces at ${q_{EP}}$ that strongly resembles the presence of ERs and change in the Fermi surface. The circuit parameters can be switched on/off repeatedly at will to realize the periodic driving.

	\end{widetext}
	\end{document}